\documentclass[12pt,a4paper]{article}
\usepackage[T1]{fontenc}
\usepackage[latin1]{inputenc}
\usepackage{amsmath}

\makeatletter

\providecommand{\LyX}{L\kern-.1667em\lower.25em\hbox{Y}\kern-.125emX\@}
\newenvironment{LyXParagraphIndent}[1]%
{
  \begin{list}{}{%
    \setlength\topsep{0pt}%
    \addtolength{\leftmargin}{#1}
    \setlength\parsep{0pt plus 1pt}%
  }
  \item[]
}
{\end{list}}

\newfont{\mcal}{rsfs10 scaled 1200}
\renewcommand{\atop}[2]{\genfrac{}{}{0pt}{}{#1}{#2}}
\usepackage{amssymb}
\usepackage{epsfig}
\newcommand{\sh}{\operatorname{sh}}
\newcommand{\ch}{\operatorname{ch}}
\renewcommand{\th}{\operatorname{th}}
\newfont{\bit}{cmbxti10 scaled 1728}

\makeatother

\begin{document}

{\par\centering {\LARGE Generalized asymptotic structure} \par}

{\par\centering {\LARGE of the ultrarelativistic} \par}

{\par\centering {\LARGE Schwarzschild black hole}\LARGE \par}

\vspace{2cm}
{\par\centering {\large Peter C. AICHELBURG\footnote[1]{email: pcaich@doppler.thp.univie.ac.at}\( ^{\dagger } \)}\large \par}
\vspace{0.3cm}

{\par\centering \textit{\small Institut für Theoretische Physik, Universität
Wien, }\small \par}

{\par\centering \textit{\small Boltzmanngasse 5, 1090 Wien, }\small \par}

{\par\centering \textit{\small AUSTRIA}\small \par}

\vspace{0.5cm}
{\par\centering {\large and }\large \par}
\vspace{0.5cm}

{\par\centering {\large Herbert BALASIN\footnote[4]{email: hbalasin@tph.tuwien.ac.at}\footnote[2]{supported by the Austrian science foundation: project P13007-PHY}}\large \par}
\vspace{0.3cm}

{\par\centering \textit{\small Institut für Theoretische Physik, TU-Wien, }\small \par}

{\par\centering \textit{\small Wiedner Hauptstraße 8-10, 1040 Wien, }\small \par}

{\par\centering \textit{\small AUSTRIA}\small \par}
\vspace{1cm}

\begin{abstract}
We discuss the asymptotic structure of the ultrarelativistic Schwarz\-schild
black hole. An explicit construction for a conformal boundary both at spatial
and null infinity is given together with the corresponding expressions for the
ADM and Bondi four-momenta.
\vspace{1cm}

\vspace{2cm}
\hspace*{8.5cm}TUW-99-28

\hspace*{8.5cm}UWthPh-1999-78
\end{abstract}

\newpage
\subsection*{\large\bit Introduction}

In a previous short comment the authors \cite{AiBa4} argued that it is possible
to assign ADM as well as Bondi four-momentum to the geometry of an ultrarelativistic
Schwarzschild black hole, which is described by the so-called Aichelburg-Sexl
metric (AS). This result is in agreement with the physically motivated expectations:
the Bondi momentum is lightlike and equal to the ADM momentum up to the instant
of retarded time when both, particle and radiation, escape to infinity and drops
to zero thereafter, leaving flat space behind. The naive calculation made use
of the standard limiting procedure to spatial and null infinity of global quantities
defined with respect to manifestly asymptotic flat coordinates. Although it
lead to the intuitively correct results the asymptotic structure of the AS metric
was not discussed. 

The goal of the present work is somewhat more ambitious and aims at the construction
of asymptopia, \( i^{0} \) as well as \( \text {\mcal I} \), for the AS-metric
. Due to the distributional nature of the geometry, which belongs to the class
of impulsive gravitational waves, the usual techniques do not immediately apply.
Therefore, to disentangle geometrical and distributional effects we also consider
the analogous electromagnetic situation in Minkowski space. The electromagnetic
field of an ultrarelativistic charge has a distributional structure similar
to that of the impulsive gravitational wave. In the first part of section two
(em-pulse), we show that the intergrals of the field at spatial and null infinity
encode the correct total charge. 

To exhibit our approach we first consider in section one the unboosted situation
and construct the asymptotic Coulomb and Schwarzschild fields. This leads, of
course, to well-known results but allows us to introduce notation and the techniques
we wish to apply: For spatial infinity we essentially follow the Ashtekar-Hansen
\cite{Ash,AshHa} approach whereas for null infinity we make use of the Penrose-like
definitions, however in terms of tensor quantities. For the construction of
the conformal boundary of the Schwarzschild field we use the flat part of the
Kerr-Schild decomposition. This will serve as a starting point for the extension
to the impulsive situation.

The main part of this paper can be found in second part of section two (AS-geometry)
where we apply the formalism to the boosted Schwarzschild metric. These calculations
not only show that it makes sense to construct a conformal boundary for this
spacetime, although in a somewhat generalized form, but also confirm our previous
results of total four-momentum.

Section zero discusses the standard form of the AS-metric and gives a brief
introduction to the concepts of direction-dependent limits and its distributional
extension.

We would like to emphasize that nowhere we make use of non-linear techniques
for generalized functions such as Colombeau theory. All quantities are well
defined in the setting of classical distributional theory.

\subsection*{\large\bit 0) Preliminaries}

It is well known that the metric 
\begin{eqnarray}
ds^{2}=-dt^{2}+dz^{2}+dx^{2}+dy^{2}+f(dt-dz)^{2}, &  & \nonumber \label{AS-metric} \\
f=-8\mu \delta (t-z)\log \sqrt{x^{2}+y^{2}} &  & \label{AS-metric} 
\end{eqnarray}
can be obtained from the Schwarzschild geometry by applying a singular boost
(or light-like contraction ) i.e. taking the limit \( \gamma \to \infty  \)
and at the same time the mass \( m\to 0 \) such that \( m\gamma =\mu =const. \)
\\
Geometrically this metric is defined by

\begin{enumerate}
\item the existence of a covariantly constant vector field, 
\item satisfying Einsteins vacuum equations (except on a null line), 
\item a curvature tensor which is completely concentrated on a null hyperplane and
that
\item its symmetry group acts on the null hyperplanes and has the structure \( \mathbb R\times ISO(2) \)
\end{enumerate}
The first two conditions imply that the metric belongs to the class of pp-waves
which in turn guarantees the existence of adapted coordinates such that the
metric takes the above form with the a general profile \( f=f(t-z,x,y) \) being
only restricted by \( (\partial ^{2}_{x}+\partial ^{2}_{y})f=0 \). 

So far the coordinates are only determined up to a (still infinite-dimen\-sional)
subgroup of diffeomorphism group, namely the so-called ``normal form preserving''
transformations. In a classical work \cite{JEK} pp-waves have been characterized
by listing all possible symmetry groups together with a corresponding canonical
representative. This classification has been extended in \cite{AiBa2,AiBa3}
to encompass also distributional wave-profiles. It follows that for impulsive
pp-waves the normal form preserving diffeomorphisms are reduced to a seven-dimensional
Lie group. The last condition then determines the form of the metric (\ref{AS-metric})
up to a rescaling of the null vector field \( p^{a}=\partial ^{a}_{t}+\partial ^{a}_{z}\to \lambda p^{a} \),
which in turn rescales \( \mu  \). If however one insists that the metric is
obtained from Schwarzschild in the ultrarelativistic limit then \( \mu  \)
is fixed by the value of \( m \).

Since the AS-geometry is distributional in nature and we are interested in its
asymptotic structure, we have to deal with asymptotic limits of distributional
quantities. Therefore we start with an informal account of direction-dependent
limits and by means of a simple example give the extension of this notion to
generalized functions (distributions). Let us begin with the (Euclidianized)
notion of direction dependent limits. Direction dependent limits are taken with
respect to a definite point \( p \) of \( M \) and its corresponding tangent
space \( T_{p}M \). It is therefore useful to first state our coordinate-conventions:
we consider a chart containing \( p\in U\subset M \) such that the local coordinates
\( (x^{1},\dots ,x^{n}) \) of \( p \) coincide with the origin. Since \( T_{p}M \)
consists of all (equivalence classes) of curves through \( p \), it can conveniently
be represented by \( x^{i}=\lambda X^{i} \), where \( X^{i} \) denote the
components of the tangent vector with respect to the coordinate basis \( \partial _{i} \),
thereby emphasizing their role as coordinates of \( T_{p}M \), when the latter
is considered as a manifold in its own right. Due to the linear structure of
\( T_{p}M \) we may identify \( \partial _{X^{i}} \), with \( \partial _{i} \).
The direction dependent limit of a scalar function \( f \) is then defined
by
\[
\lim _{\lambda \to 0}f(x(\lambda ))=:F(X)\qquad X=\dot{x}(0),\]
provided it only depends on the tangent vector. The limit (if it exists) does
actually only depend on the projective class of \( X \). If the manifold is
equipped with a (Riemannian) metric it may be used to interpret \( F \) as
a function on the (unit) \( S^{n-1} \) in \( T_{p}M \) by requiring \( X^{2}=1. \)
This approach has been successfully applied (in the Lorentzian context) to investigate
the asymptotic structure of spacetime \cite{Ash,AshHa}.

In order to generalize this notion to allow for generalized function consider
the following example: \( \mathbb R^{2} \) equipped with the standard Euclidean
metric. The asymptotic limit of a function \( F:\mathbb R^{2}\to \mathbb R \)
is calculated by evaluating 
\[
\lim _{\rho \to \infty }F(\rho \cos \phi ,\rho \sin \phi )=:f(\phi ).\]
This limit (if it exists) will in general depend on the choice of direction
along which we approach infinity. In a more geometrical picture we may think
of \( \mathbb R^{2} \) as being compactified to \( S^{2} \) and interpret
this limit as a function \( f \) on the unit-\( S^{1} \) in the tangent space
at the ``north pole'' of \( S^{2} \) representing the infinitely distant
point of \( \mathbb R^{2} \). Making use of test forms \( \tilde{\varphi }\in \Omega _{0}^{\infty }(S^{1}) \)\footnote[2]{Here and in the following distributions will be regarded as linear functionals acting on \( C^{\infty } \) test \( n \)-forms of compact support rather than test functions. This point of view is the natural one for generalized scalar functions, since the corresponding regular functionals are generated via integration of a scalar function against an \( n \)-form.}
defined with respect to this \( S^{1} \), the limit gives rise to a regular
functional \( f \) via
\[
\lim _{\rho \to \infty }\int F(\rho \cos \phi ,\rho \sin \phi )\varphi (\phi )d\phi =:(f,\tilde{\varphi }).\]
In order to see how this definition extends to singular functionals let us raise
the question what the asymptotic limit of \( \delta (y) \) -- a genuine distribution
on \( \mathbb R^{2} \) in global Cartesian \( (x,y) \) coordinates -- is.
At first glance it may seem that this limit does not exist since \( \delta (y) \)
has the same (infinite) value at its support \( y=0 \). However, taking a closer
look and using the above prescription for regular functionals we find
\[
\lim _{\rho \to \infty }(\delta (y),\tilde{\varphi })=\lim _{\rho \to \infty }(\delta (\rho \sin \phi ),\tilde{\varphi })=\lim _{\rho \to \infty }\frac{1}{\rho }(\varphi (0)+\varphi (\pi ))=0.\]
This result is not as surprising as it might look in the first place. It exhibits
the negative homogeneity of the delta function as the reason for the vanishing
of the limiting distribution. It is precisely this behavior which will turn
out to be crucial for the existence of the asymptotic quantities in the following
sections.

\section*{\large\bit 1) Coulomb charge and Schwarzschild \\
\hspace*{9cm}four-momentum}

\subsection*{\bit Coulomb }

Consider the Coulomb potential and its field strength in Minkowski space \( M \)
\begin{eqnarray*}
A & = & -\frac{e}{r}dt,\\
F & = & dA=\frac{e}{r^{2}}dr\wedge dt\hspace {2cm}\atop {u=t-r}{v=t+r}\\
 & = & \frac{2e}{(v-u)^{2}}dv\wedge du.
\end{eqnarray*}
 \( \bullet  \) \( i^{0} \) and ADM-charge: In order to evaluate the (total)
ADM charge let us switch to a regular coordinate system at (spatial) infinity
\[
\atop {\bar{v}=-1/u}{\bar{u}=-1/v}\qquad \atop {d\bar{v}=(1/u^{2})du}{d\bar{u}=(1/v^{2})dv}\]
With respect to these coordinates \( F \) becomes
\[
F=\frac{2e}{\left( -1/\bar{u}+1/\bar{v}\right) ^{2}}\frac{1}{\bar{u}^{2}}\frac{1}{\bar{v}^{2}}d\bar{u}\wedge d\bar{v}=\frac{2e}{\left( \bar{v}-\bar{u}\right) ^{2}}d\bar{u}\wedge d\bar{v},\]
and the metric
\[
ds^{2}=-dudv+\frac{(v-u)^{2}}{4}d\Omega ^{2}=\frac{1}{\bar{u}^{2}\bar{v}^{2}}(-d\bar{u}d\bar{v}+\frac{(\bar{v}-\bar{u})^{2}}{4}d\Omega ^{2})\qquad \Omega =-\bar{u}\bar{v}.\]
Following \cite{Ash,AshHa} we take the direction dependent limit of \( F \)
using \( \bar{u}=\lambda U,\bar{v}=\lambda V \) and \( \lambda \to 0 \) and
find
\[
\lim _{\lambda \to 0}(\Omega F)=(-\lambda ^{2}UV)\frac{2e}{\lambda ^{2}(V-U)^{2}}dU\wedge dV=(-UV)\frac{2e}{(V-U)^{2}}dU\wedge dV,\]
which is a tensor-field in \( T_{i^{0}}\hat{M} \), where \( \hat{M} \) refers
to the standard conformal compactification of \( M \). Physically only the
(spatial) projective information is relevant. This information is most conveniently
encoded in the hyperboloid \( \mathcal{D}: \) \( UV=-1 \) and the field-strength
tensor at \( \mathcal{D} \)
\[
\mathbb {F}:=\lim _{\lambda \to 0}(\Omega F)|_{\mathcal{D}}=\frac{2e}{(V+1/V)^{2}}dU\wedge dV,\]
which is a section of the pull-back bundle of the tangent bundle of \( T_{i^{0}}\hat{M} \)
to \( \mathcal{D} \) . Using the normal \( \eta ^{a}=U\partial _{U}^{a}+V\partial _{V}^{a} \)
of \( \mathcal{D} \) the flux 2-form \( \frac{1}{2}{}^{4}\epsilon _{abcd}\mathbb {F}^{ab} \)
may be decomposed
\[
\frac{1}{2}{}^{4}\epsilon _{abcd}\mathbb {F}^{ab}=-(\eta _{a}\mathbb {F}^{ab}){}^{3}\epsilon _{bcd}-\eta _{[c}{}^{3}\epsilon _{abd]}\mathbb {F}^{ab},\qquad {}^{4}\epsilon _{abcd}=-\eta _{[a}{}^{3}\epsilon _{bcd]}.\]
Integration over a 2-sphere section of \( \mathcal{D} \) gives the total or
``ADM'' charge

\begin{eqnarray*}
Q^{ADM}=-\frac{1}{4\pi }\int _{S^{2}}\frac{1}{2}{}^{4}\epsilon _{abcd}\mathbb {F}^{ab}=\frac{1}{4\pi }\int _{S^{2}}{}^{3}\epsilon _{abc}E^{a}=e &  & \\
E:=\eta \rfloor \mathbb {F}=-\frac{4e}{(V+1/V)^{2}}\frac{dV}{V}=-\frac{e}{\ch ^{2}\tau }d\tau ,\quad \{V=e^{\tau }\} &  & 
\end{eqnarray*}
where the second integral expression is the usual Ashtekar expression for the
ADM-charge.\\
\( \bullet  \) \( \text {\mcal I}^{+} \) and Bondi-charge: As for the ADM-charge
we begin with the field strength \( F \) and pass to coordinates regular in
a neighborhood of \( \text {\mcal I}^{+} \)
\begin{eqnarray*}
F & = & \frac{2e}{(v-u)^{2}}dv\wedge du\qquad v=\frac{1}{\bar{v}}\\
 & = & \frac{2e}{(1-u\bar{v})^{2}}\frac{1}{v^{2}}\left( -\frac{d\bar{v}}{\bar{v}^{2}}\right) \wedge du\\
 & = & \frac{2e}{(1-u\bar{v})^{2}}du\wedge d\bar{v}.
\end{eqnarray*}
In the limit \( \bar{v}\to 0 \) this becomes
\[
\mathbb {F}:=\lim _{\bar{v}\to 0}F=2edu\wedge d\bar{v}.\]
Taking the corresponding transformation of the metric into account 
\[
ds^{2}=-dudv+\frac{(v-u)^{2}}{4}d\Omega ^{2}=\frac{1}{\bar{v}^{2}}(dud\bar{v}+\frac{(1-u\bar{v})^{2}}{4}d\Omega ^{2})\qquad \Omega =\bar{v}\]
allows us to calculate the flux 2-form and the Bondi-charge as its integral
over a 2-sphere section of \( \text {\mcal I}^{+} \)
\begin{eqnarray*}
*\mathbb {F}=*(-4e(e^{u}\wedge e^{\bar{v}}))=4e\frac{1}{4}d^{2}\Omega =ed^{2}\Omega  &  & \\
Q^{Bondi}=\frac{1}{4\pi }\int _{S^{2}}*\mathbb {F}=e, &  & 
\end{eqnarray*}
which remains constant since no charge escapes to null-infinity.

\subsection*{\bit Schwarzschild}

In the corresponding gravitational situation the Riemann (Weyl) tensor plays
the role of the field strength \( F_{ab} \). Using adapted coordinates for
the Kerr-Schild decomposition of Schwarzschild
\[
g_{ab}=\eta _{ab}+fk_{a}k_{b}\qquad f=\frac{2m}{r},\, \, k^{a}=\partial _{t}^{a}+\partial _{r}^{a},\]
the mixed form of the Riemann-tensor \( R^{ab}\, _{cd} \) (\( \mathbf{Riemann} \))
becomes

\begin{eqnarray*}
\mathbf{Riemann} & = & \frac{f''}{2}(\partial _{t}\wedge \partial _{r})(dt\wedge dr)\\
 &  & +\frac{f'}{2r}\left( (\partial _{t}\wedge \tilde{E}_{i})(dt\wedge \tilde{e}^{i})+(\partial _{r}\wedge \tilde{E}_{i})(dr\wedge \tilde{e}^{i})\right) \\
 &  & +\frac{f}{2r^{2}}(\tilde{E}_{i}\wedge \tilde{E}_{j})(\tilde{e}^{i}\wedge \tilde{e}^{j})=\{\atop {u=t-r}{v=t+r}\}\\
 & = & \frac{f''}{2}(\partial _{u}\wedge \partial _{v})(du\wedge dv)\\
 &  & +\frac{f'}{2r}\left( (\partial _{u}\wedge \tilde{E}_{i})(du\wedge \tilde{e}^{i})+(\partial _{v}\wedge \tilde{E}_{i})(dv\wedge \tilde{e}^{i})\right) \\
 &  & +\frac{f}{2r^{2}}(\tilde{E}_{i}\wedge \tilde{E}_{j})(\tilde{e}^{i}\wedge \tilde{e}^{j}),
\end{eqnarray*}
where \( \tilde{E}_{i},\, \tilde{e}^{i} \) refer to an orthonormal dyad and
co-dyad for the unit two-sphere respectively. \\
\( \bullet  \) \( i^{0} \) and ADM-momentum: Changing coordinates to a patch
that is regular at spatial infinity 
\[
\atop {\bar{u}=-1/v}{\bar{v}=-1/u}\quad \atop {\partial _{u}=\bar{u}^{2}\partial _{\bar{u}}}{\partial _{v}=\bar{v}^{2}\partial _{\bar{v}}}\quad \atop {du=1/\bar{u}^{2}d\bar{u}}{dv=1/\bar{v}^{2}d\bar{v}}\]
 the expression for \( \mathbf{Riemann} \) becomes
\begin{eqnarray*}
\mathbf{Riemann} & = & \frac{f''}{2}(\partial _{\bar{u}}\wedge \partial _{\bar{v}})(d\bar{u}\wedge d\bar{v})+\\
 &  & +\frac{f'}{2r}\left( (\partial _{\bar{u}}\wedge \tilde{E}_{i})(d\bar{u}\wedge \tilde{e}^{i})+(\partial _{\bar{v}}\wedge \tilde{E}_{i})(d\bar{v}\wedge \tilde{e}^{i})\right) \\
 &  & +\frac{f}{2r^{2}}(\tilde{E}_{i}\wedge \tilde{E}_{j})(\tilde{e}^{i}\wedge \tilde{e}^{j}).
\end{eqnarray*}
Taking the direction dependent limit by using \( \bar{u}=\lambda U,\, \bar{v}=\lambda V \)
and \( \lambda \to 0 \) we find
\begin{eqnarray*}
\lim _{\lambda \to 0}(\Omega ^{-3/2}\mathbf{Riemann}) & = & \frac{16M(-UV)^{3/2}}{(V-U)^{3}}(\partial _{U}\wedge \partial _{V})(dU\wedge dV)\\
 &  & \hspace {-2cm}-\frac{8M(-UV)^{3/2}}{(V-U)^{3}}\left( (\partial _{U}\wedge \tilde{E}_{i})(dU\wedge \tilde{e}^{i})+(\partial _{V}\wedge \tilde{E}_{i})(dV\wedge \tilde{e}^{i})\right) \\
 &  & \hspace {-2cm}+\frac{8M(-UV)^{3/2}}{(V-U)^{3}}(\tilde{E}_{i}\wedge \tilde{E}_{j})(\tilde{e}^{i}\wedge \tilde{e}^{j}).
\end{eqnarray*}
Since, as in the electromagnetic case, only the (spatially) projective information
is physically relevant we will use the asymptotic limit of the Riemann tensor
at the hyperboloid \( \mathcal{D}: \) \( UV=-1 \)
\begin{eqnarray*}
\mathbb {R}^{ab}\, _{cd} & := & \lim _{\lambda \to 0}(\Omega ^{-3/2}R^{ab}\, _{cd})|_{\mathcal{D}}=\frac{16M}{(V+1/V)^{3}}(\partial _{U}\wedge \partial _{V})^{ab}(dU\wedge dV)_{cd}\\
 &  & -\frac{8M}{(V+1/V)^{3}}\left( (\partial _{U}\wedge \tilde{E}_{i})^{ab}(dU\wedge \tilde{e}^{i})_{cd}+(\partial _{V}\wedge \tilde{E}_{i})^{ab}(dV\wedge \tilde{e}^{i})_{cd}\right) \\
 &  & \frac{8M}{(V+1/V)^{3}}(\tilde{E}_{i}\wedge \tilde{E}_{j})^{ab}(\tilde{e}^{i}\wedge \tilde{e}^{j})_{cd}.
\end{eqnarray*}
The asymptotic form of the bivector potential is obtained from
\begin{eqnarray*}
\lim _{\lambda \to 0}\Omega ^{-3/2}(x\wedge \alpha ) & = & -(\alpha ^{t}-(\alpha \cdot e_{r}))\frac{UV^{2}}{(-UV)^{3/2}}\partial _{U}\wedge \partial _{V}\\
 &  & \hspace {-3cm}+(\alpha ^{t}+(\alpha \cdot e_{r}))\frac{U^{2}V}{(-UV)^{3/2}}\partial _{U}\wedge \partial _{V}+\frac{UV^{2}}{(-UV)^{3/2}}\frac{2(\alpha \cdot e_{\theta })}{V-U}\partial _{V}\wedge \partial _{\theta }\\
 &  & \hspace {-3cm}+\frac{U^{2}V}{(-UV)^{3/2}}\frac{2(\alpha \cdot e_{\theta })}{(V-U)}\partial _{U}\wedge \partial _{\theta }+\frac{UV^{2}}{(-UV)^{3/2}}\frac{2(\alpha \cdot e_{\phi })}{(V-U)\sin \theta }\partial _{V}\wedge \partial _{\phi }\\
 &  & \hspace {-3cm}+\frac{U^{2}}{(-UV)^{3/2}}\frac{2(\alpha \cdot e_{\phi })}{(V-U)\sin \theta }\partial _{U}\wedge \partial _{\phi }
\end{eqnarray*}
where \( \alpha  \) denotes a translation with respect to flat part of the
Kerr-Schild decomposition. The expression for the bivector potential at \( \mathcal{D} \)
becomes 
\begin{eqnarray*}
Q:=\lim _{\lambda \to 0}\Omega ^{-3/2}(x\wedge \alpha )|_{\mathcal{D}} & = & (\alpha ^{t}-(\alpha \cdot e_{r}))V\partial _{U}\wedge \partial _{V}\\
 &  & \hspace {-4cm}+(\alpha ^{t}+(\alpha \cdot e_{r}))\frac{1}{V}\partial _{U}\wedge \partial _{V}-\frac{2(\alpha \cdot e_{\theta })}{V+1/V}V\partial _{V}\wedge \partial _{\theta }+\frac{2(\alpha \cdot e_{\theta })}{V+1/V}\frac{1}{V}\partial _{U}\wedge \partial _{\theta }\\
 &  & \hspace {-4cm}-\frac{2(\alpha \cdot e_{\phi })}{(V+1/V)\sin \theta }V\partial _{V}\wedge \partial _{\phi }+\frac{2(\alpha \cdot e_{\phi })}{(V+1/V)\sin \theta }\frac{1}{V}\partial _{U}\wedge \partial _{\phi }.
\end{eqnarray*}
 Contraction of \( \mathbb {R}^{ab}\, _{cd} \) with \( Q^{ab} \) yields the
gravitational field-strength. Using the normal \( \eta ^{a}=U\partial _{U}^{a}+V\partial _{V}^{a} \)
of \( \mathcal{D} \) together with the expression for an asymptotic translation
\begin{eqnarray*}
\lim _{\lambda \to 0}(\Omega ^{-1}\alpha ^{a})|_{\mathcal{D}} & = & (\alpha ^{t}-(\alpha \cdot e_{r}))V^{2}\partial _{V}^{a}+(\alpha ^{t}+(\alpha \cdot e_{r}))\frac{1}{V^{2}}\partial _{U}^{a}\\
 &  & +(\alpha \cdot e_{\theta })\frac{2}{V+1/V}\partial _{\theta }^{a}+(\alpha \cdot e_{\phi })\frac{2}{V+1/V}\partial ^{a}_{\phi })=\\
 & = & (\alpha ^{t}\ch \tau -(\alpha \cdot e_{r})\sh \tau )\partial _{\tau }^{a}+\\
 &  & (\alpha \cdot e_{\theta })\frac{1}{\ch \tau }\partial _{\theta }^{a}+(\alpha \cdot e_{\phi })\frac{1}{\ch \tau \sin \theta }\partial _{\phi }^{a}
\end{eqnarray*}
and taking into account that \( Q^{ab}=-(\eta \wedge \alpha )^{ab} \)\footnote{%
where by a slight abuse of notation with have used the same symbol \( \alpha  \)
for both the Spi-translation and the translation relative to \( \eta _{ab} \)
} , the corresponding flux 2-form may be decomposed as
\[
\frac{1}{4}{}^{4}\epsilon _{abcd}\mathbb {R}^{ab}\, _{mn}Q^{mn}={}^{3}\epsilon _{bcd}(\eta _{a}\mathbb {R}^{ab}\, _{mn}\eta ^{m})\alpha ^{n}+\eta _{[c}\, {}^{3}\epsilon _{d]ab}\mathbb {R}^{ab}\, _{mn}\eta ^{m}\alpha ^{n}.\]
 Integration of the flux 2-form over an \( S^{2} \)-section of \( \mathcal{D} \)
gives the ADM-momentum 
\[
P^{ADM}\cdot \alpha =-\frac{1}{8\pi }\int _{S^{2}}\frac{1}{4}{}^{4}\epsilon _{abcd}(\mathbb {R}^{ab}\, _{mn}Q^{cd})=-\frac{1}{8\pi }\int _{S^{2}}{}^{3}\epsilon _{bcd}(E^{b}\, _{n}\alpha ^{n}),\]
where the Ashtekar-Hansen gravi-electric field is defined by
\begin{eqnarray*}
E^{a}\, _{b}:=\eta _{c}\eta ^{d}\mathbb {R}^{ca}\, _{db} & = & \frac{8M}{(V+1/V)^{3}}(\frac{1}{V}\partial _{U}+V\partial _{V})^{a}(\frac{1}{V}dV+VdU)_{b}\\
 &  & -\frac{8M}{(V+1/V)^{3}}\tilde{E}^{a}_{i}\tilde{e}_{b}^{i}\\
 & = & \frac{M}{\ch ^{3}\tau }(2\partial _{\tau }^{a}d\tau _{b}-\tilde{E}^{a}_{i}\tilde{e}^{i}_{b})\qquad \{V=e^{\tau }\}.
\end{eqnarray*}
 Evaluating the contraction with an asymptotic translation

\[
E^{a}\, _{b}\alpha ^{b}=\frac{M}{\ch ^{3}\tau }\left( 2(\alpha ^{t}\ch \tau -(\alpha \cdot e_{_{r}})\sh \tau )\partial _{\tau }^{a}-\frac{(\alpha \cdot e_{\theta })}{\ch \tau }\partial _{\theta }^{a}-\frac{(\alpha \cdot e_{\phi })}{\ch \tau \sin \theta }\partial _{\phi }^{a}\right) ,\]
where \( e_{r},e_{\theta },e_{\phi } \) denote the unit vectors with respect
to standard spherical polar coordinates, over a \( \tau =const \) section of
\( \mathcal{D} \) finally gives the ADM four-momentum

\[
P^{ADM}\cdot \alpha =-\frac{1}{8\pi }\int _{S^{2}_{\tau }}{}^{3}\epsilon _{abc}(E^{a}\, _{m}\alpha ^{m})=-M\alpha ^{t}.\]
  \( \bullet  \) \( \text {\mcal I}^{+} \) and Bondi four-momentum: Relative
to coordinates which are regular in a neighborhood of \( \text {\mcal I}^{+} \)
\[
\bar{v}=\frac{1}{v}\quad \partial _{v}=-\bar{v}^{2}\partial _{\bar{v}}\qquad dv=-\frac{1}{\bar{v}^{2}}d\bar{v}\]
the expression for the Riemann-tensor becomes
\begin{eqnarray*}
\mathbf{Riemann} & = & \frac{f''}{2}(\partial _{u}\wedge \partial _{\bar{v}})(du\wedge d\bar{v})\\
 &  & +\frac{f'}{2r}\left( (\partial _{u}\wedge \tilde{E}_{i})(du\wedge \tilde{e}^{i})+(\partial _{\bar{v}}\wedge \tilde{E}_{i})(d\bar{v}\wedge \tilde{e}^{i})\right) \\
 &  & +\frac{f}{2r^{2}}(\tilde{E}_{i}\wedge \tilde{E}_{j})(\tilde{e}^{i}\wedge \tilde{e}^{j}).
\end{eqnarray*}
Taking the limit to \( \text {\mcal I}^{+} \) we obtain for the rescaled Riemann
(Weyl) tensor
\begin{eqnarray*}
\mathbb {R}^{ab}\, _{cd} & = & \lim _{\bar{v}\to 0}(\Omega ^{-3}R^{ab}\, _{cd})=16M(\partial _{u}\wedge \partial _{\bar{v}})^{ab}(du\wedge d\bar{v})_{cd}\\
 &  & -8M\left( (\partial _{u}\wedge \tilde{E}_{i})^{ab}(du\wedge \tilde{e}^{i})_{cd}+(\partial _{\bar{v}}\wedge \tilde{E}_{i})^{ab}(d\bar{v}\wedge \tilde{e}^{i})_{cd}\right) \\
 &  & +8M(\tilde{E}_{i}\wedge \tilde{E}_{j})^{ab}(\tilde{e}^{i}\wedge \tilde{e}^{j})_{cd},
\end{eqnarray*}
and the bivector potential
\begin{eqnarray*}
Q^{ab}=\lim _{\bar{v}\to 0}(\Omega ^{-1}(x\wedge \alpha )^{ab}) & = & (\alpha ^{t}-(\alpha \cdot e_{r}))(\partial _{u}\wedge \partial _{\bar{v}})^{ab}\\
 &  & +2u(\alpha \cdot e_{\theta })(\partial _{u}\wedge \partial _{\theta })^{ab}+2u\frac{(\alpha \cdot e_{\phi })}{\sin \theta }(\partial _{u}\wedge \partial _{\phi })^{ab}.
\end{eqnarray*}
Contracting \( \mathbb {R}^{ab}\, _{cd} \) with \( Q^{ab} \) 
\begin{eqnarray*}
\frac{1}{2}\mathbb {R}^{ab}\, _{cd}Q^{cd} & = & 16M(\alpha ^{t}-(\alpha \cdot e_{r}))\partial _{u}\wedge \partial _{\bar{v}}\\
 &  & -16M(\alpha \cdot e_{\theta })u\partial _{u}\wedge \partial _{\theta }-16M\frac{(\alpha \cdot e_{\phi })}{\sin \theta }u\partial _{u}\wedge \partial _{\phi ,}
\end{eqnarray*}
once again yields the ``gravitational'' field-strength bivector whose flux
through an \( S^{2}_{u} \) section of \( \text {\mcal I}^{+} \) gives the
Bondi four-momentum
\begin{eqnarray*}
\frac{1}{4}\, ^{4}\epsilon _{abcd}\mathbb {R}^{ab}\, _{mn}Q^{mn}=-2M(\alpha ^{t}-(\alpha \cdot e_{r}))d^{2}\Omega _{cd} &  & \\
P^{Bondi}\cdot \alpha =\frac{1}{8\pi }\int _{S^{2}_{u}}\, ^{4}\epsilon _{abcd}\mathbb {R}^{ab}\, _{mn}Q^{mn}\frac{1}{4}=-M\alpha ^{t}, &  & 
\end{eqnarray*}
which coincides with the ADM momentum since there is no gravitational radiation
in the Schwarzschild spacetime. In summary we have shown that the ``Kerr-Schild
compactification'', i.e. constructing the conformal boundary from the flat
part of the Kerr-Schild decomposition yields the correct Bondi and ADM four-momenta.

\section*{\large\bit 2) EM-pulse and AS geometry}

\subsection*{\bit EM-pulse}

The ultrarelativistic limit of the Coulomb field, i.e. the electromagnetic field
of a point charge moving at the speed of light, is most easily constructed by
boosting the corresponding four-current. Relative to the rest frame of the charge
we find
\[
j^{a}=e\delta ^{(3)}(x)\partial ^{a}_{t}=e\delta (Q\cdot x)\delta ^{(2)}(x)P^{a}\qquad P^{a}=m\partial _{t}^{a},\, Q^{a}=m\partial _{z}^{a},\]
where \( P^{a} \) and \( Q^{a} \) represent the respectively time and spacelike
vectors spanning the boost plane, which are conveniently normalized in order
to allow a null limit. Denoting their common null limit by \( p^{a} \) the
EM-pulse current becomes
\[
j^{a}=e\delta (px)\delta ^{(2)}(x)p^{a}\]
which in turn via Maxwell's equations gives rise to the EM-pulse potential and
field-strength
\begin{eqnarray*}
A & = & -2e\delta (px)\log \rho \, pdx,\hspace {5cm}\rho ^{2}=x^{2}+y^{2}\\
F & = & -2e\delta (px)\frac{1}{\rho }d\rho \wedge pdx\hspace {4.9cm}\atop {u=t-r}{v=t+r}\\
 & = & -\frac{2e\delta (u(1+\cos \theta )+v(1-\cos \theta ))}{(v-u)\sin \theta }(du(1+\cos \theta )+dv(1-\cos \theta )\\
 &  & +(v-u)\sin \theta d\theta )\wedge ((dv-du)\sin \theta +(v-u)\cos \theta d\theta )\\
 & = & -\frac{2e\delta (u(1+\cos \theta )+v(1-\cos \theta ))}{(v-u)\sin \theta }(2du\wedge dv\sin \theta +\\
 &  & +du\wedge d\theta (1+\cos \theta )(v-u)-dv\wedge d\theta (1-\cos \theta )(v-u)).
\end{eqnarray*}
Note that field is completely concentrated on the null-hyperplane \( px=-t+z=0 \)
and has a singularity along the generator \( \rho =0, \) which represents the
support of the current of the boosted charge. Moreover, its character is completely
null, i.e. \( F\wedge F \) as well as \( F\wedge *F \) vanish. 

\vspace{0.5cm}

\begin{picture}(0,0)%
\epsfig{file=zelt2.pstex}%
\end{picture}%
\setlength{\unitlength}{3947sp}%
\begingroup\makeatletter\ifx\SetFigFont\undefined%
\gdef\SetFigFont#1#2#3#4#5{%
  \reset@font\fontsize{#1}{#2pt}%
  \fontfamily{#3}\fontseries{#4}\fontshape{#5}%
  \selectfont}%
\fi\endgroup%
\begin{picture}(5062,4795)(1689,-5294)
\put(4351,-5236){\makebox(0,0)[lb]{\smash{\SetFigFont{12}{14.4}{\familydefault}{\mddefault}{\updefault}$\{px=0\}\cap\{\phi=const.\}$}}}
\put(5851,-1636){\makebox(0,0)[lb]{\smash{\SetFigFont{12}{14.4}{\rmdefault}{\mddefault}{\updefault}$\text{\mcal I}^+$}}}
\put(6751,-2686){\makebox(0,0)[lb]{\smash{\SetFigFont{12}{14.4}{\rmdefault}{\mddefault}{\updefault}the particle}}}
\put(6751,-2461){\makebox(0,0)[lb]{\smash{\SetFigFont{12}{14.4}{\rmdefault}{\mddefault}{\updefault}worldline of}}}
\end{picture}

\begin{LyXParagraphIndent}{-0.5cm}
\vspace{0.5cm}
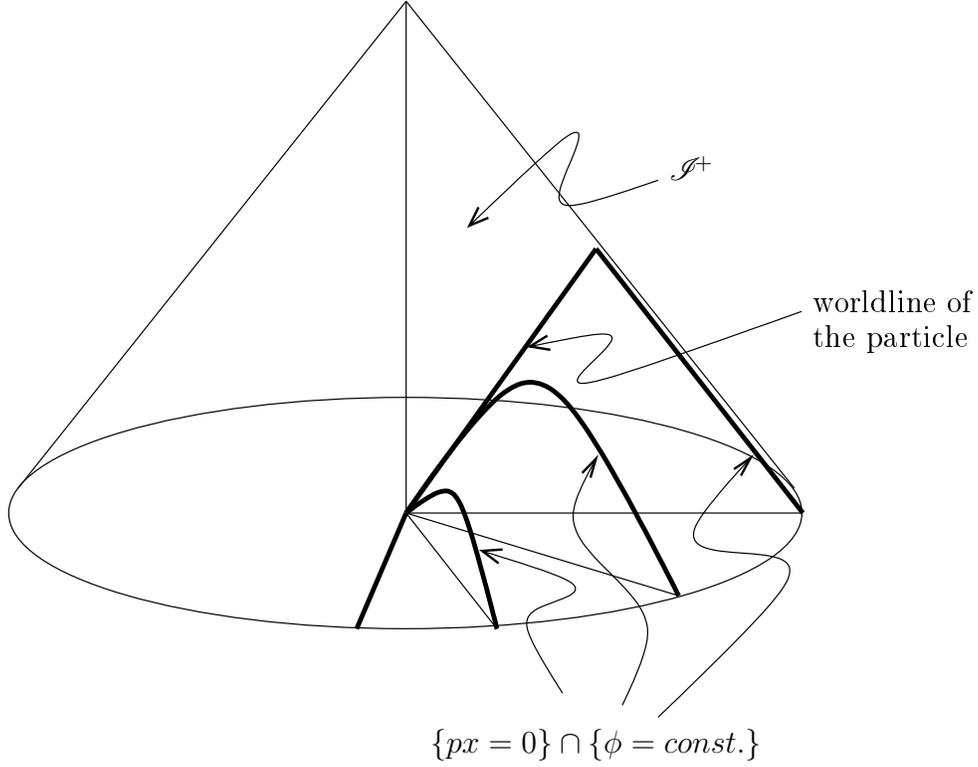
Figure 1: \( \phi =const \) sections of the pulse-plane \( px=0 \) in conformally
compactified Minkowski space
\vspace{1cm}

\end{LyXParagraphIndent}

\noindent \( \bullet  \) \( i^{0} \) and ADM-charge: Proceeding analogously
to the Coulomb situation the field strength becomes relative to coordinates
\( \bar{u}=-1/v,\bar{v}=-1/u \) regular at \( i^{0} \) 

\begin{eqnarray*}
F & = & \frac{2e\delta (\bar{v}(1+\cos \theta )+\bar{u}(1-\cos \theta ))}{(\bar{v}-\bar{u})\sin \theta }(2d\bar{u}\wedge d\bar{v}\sin \theta +\\
 &  & \hspace {1cm}+\frac{\bar{v}}{\bar{u}}d\bar{u}\wedge d\theta (1+\cos \theta )(\bar{v}-\bar{u})-\frac{\bar{u}}{\bar{v}}d\bar{v}\wedge d\theta (1-\cos \theta )(\bar{v}-\bar{u}))
\end{eqnarray*}
 Taking the direction dependent of \( F \) limit to \( i^{0} \) using \( \bar{u}=\lambda U \)
and \( \bar{v}=\lambda V \) we find
\begin{eqnarray*}
\lim _{\lambda \to 0}(\Omega F) & = & (-UV)\frac{2e\delta (U(1+\cos \theta )+V(1-\cos \theta ))}{(V-U)\sin \theta }(-2dU\wedge dV\sin \theta +\\
 &  & -\frac{V}{U}(1+\cos \theta )dV\wedge d\theta (V-U)+\frac{U}{V}(1-\cos \theta )dU\wedge d\theta (V-U)),
\end{eqnarray*}
which is a tensor distribution on \( T_{i^{0}}\hat{M} \). the field at infinity
is completely determined by its restriction to \( \mathcal{D}:UV=-1 \)
\begin{eqnarray*}
\mathbb {F} & := & \lim _{\lambda \to 0}(\Omega F)|_{\mathcal{D}}=\frac{2e\delta ((1/V)(1+\cos \theta )-V(1-\cos \theta ))}{(V+1/V)\sin \theta }(-2dU\wedge dV\sin \theta +\\
 &  & +V^{2}(1+\cos \theta )dV\wedge d\theta (V+1/V)-\frac{1}{V^{2}}(1-\cos \theta )dU\wedge d\theta (V+1/V)).
\end{eqnarray*}
 \vspace{0.5cm}

\hspace*{2cm}\begin{picture}(0,0)%
\epsfig{file=hyp2.pstex}%
\end{picture}%
\setlength{\unitlength}{3947sp}%
\begingroup\makeatletter\ifx\SetFigFont\undefined%
\gdef\SetFigFont#1#2#3#4#5{%
  \reset@font\fontsize{#1}{#2pt}%
  \fontfamily{#3}\fontseries{#4}\fontshape{#5}%
  \selectfont}%
\fi\endgroup%
\begin{picture}(3158,2866)(1493,-2769)
\put(4351,-436){\makebox(0,0)[lb]{\smash{\SetFigFont{12}{14.4}{\rmdefault}{\mddefault}{\updefault}support of the}}}
\put(4351,-661){\makebox(0,0)[lb]{\smash{\SetFigFont{12}{14.4}{\rmdefault}{\mddefault}{\updefault}asymptotic field }}}
\put(4651,-1561){\makebox(0,0)[lb]{\smash{\SetFigFont{12}{14.4}{\rmdefault}{\mddefault}{\updefault}$\mathcal{D}$}}}
\put(4201,-1786){\makebox(0,0)[lb]{\smash{\SetFigFont{12}{14.4}{\rmdefault}{\mddefault}{\updefault}hyperboloid at $i^0$}}}
\end{picture}

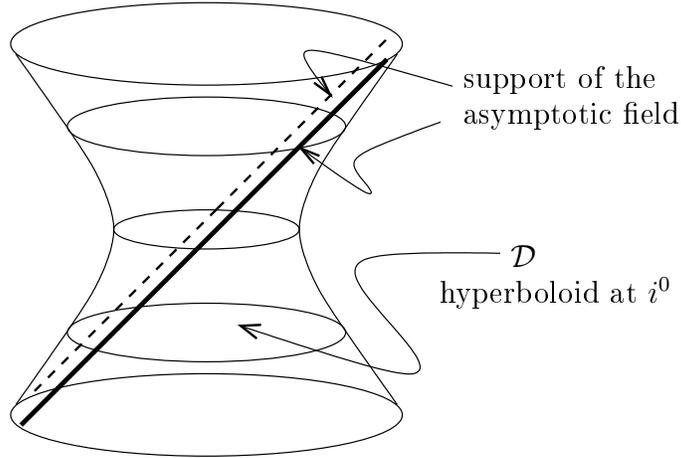
Figure 2: Support of the field strength on the set \( \mathcal{D} \) of spacelike
directions \\
~\\
~\\
Integrating the flux 2-form over an \( S^{2} \) section of \( \mathcal{D} \)
gives the ADM charge
\[
Q^{ADM}=-\frac{1}{4\pi }\int _{S^{2}}\frac{1}{2}{}^{4}\epsilon _{abcd}\mathbb {F}^{ab}=\frac{1}{4\pi }\int _{S^{2}}{}^{3}\epsilon _{abc}E^{a},\]
where the electric field in the last equation is obtained from

\begin{eqnarray*}
E & = & \eta \rfloor \mathbb {F}=2e\frac{\delta ((1/V)(1+\cos \theta )-V(1-\cos \theta ))}{(V+1/V)\sin \theta }\\
 &  & (2\sin \theta ((1/V)dV+VdU)-V(1+\cos \theta )(V+1/V)d\theta +\\
 &  & +(1/V)(1-\cos \theta )(V+1/V)d\theta )=\\
 &  & =2e\frac{\delta (\th \tau -\cos \theta )}{\ch ^{2}\tau }(d\tau +\ch \tau d\theta )=2ed\Theta (\th \tau -\cos \theta )\qquad \{V=e^{\tau }\}.
\end{eqnarray*}
The last expression explicitly exhibits the exactness of \( E \), which automatically
entails its closedness. Let us now show that it is also co-closed, i.e. \( D_{a}E^{a}=0 \).
Using the volume form together with a null triad on \( \mathcal{D} \)
\begin{eqnarray*}
 &  & \omega =p\wedge \bar{p}\wedge \ch \tau \sin \theta d\phi ,\\
 &  & p=d\tau +\ch \tau d\theta \quad \bar{p}=d\tau -\ch \tau d\theta ,\quad p^{2}=\bar{p}^{2}=0,\, \, p\cdot \bar{p}=-1,
\end{eqnarray*}
we find
\begin{eqnarray*}
*E=E\rfloor \omega =\frac{2e\delta }{\cosh ^{2}\tau }p\wedge \cosh \tau \sin \theta d\phi =\delta \, \left( \frac{d\tau }{\cosh ^{2}\tau }+\sin \theta d\theta \right) \wedge d\phi  &  & \\
=2ed\Theta (\tanh \tau -\cos \theta )\wedge d\phi \quad \Rightarrow \quad d*E=0. &  & 
\end{eqnarray*}
 Therefore its flux-integral over any (topologically \( S^{2} \) ) section
of \( \mathcal{D} \) has the same value
\[
Q^{ADM}=\frac{1}{4\pi }\int _{S_{\tau }^{2}}{}^{3}\epsilon _{abc}E^{a}=e.\]
 \( \bullet  \) \( \text {\mcal I}^{+} \) and Bondi-charge: In a neighborhood
of \( \text {\mcal I}^{+} \) we find the following expression for the electromagnetic
field-strength 
\begin{eqnarray*}
F & = & 2e\delta (u\bar{v}(1+\cos \theta )+(1-\cos \theta ))\\
 &  & (\frac{2}{1-u\bar{v}}du\wedge d\bar{v}-\bar{v}\frac{1+\cos \theta }{\sin \theta }du\wedge d\theta -\frac{1}{\bar{v}}\frac{1-\cos \theta }{\sin \theta }d\bar{v}\wedge d\theta )\\
 &  & \bar{v}=1/v.
\end{eqnarray*}
Taking into account that in the limit \( \bar{v}\to 0 \) the support of the
\( \delta  \)-function tends to \( \theta =0 \), the north pole of \( S^{2} \),
we find by using\footnote[2]{Note that the support of the $\delta$-function tends to the north-pole, where spherical polar coordinates become singular. Nevertheless using regular (stereographic) coordinates leads to the same result.}
\begin{eqnarray*}
\lim _{\bar{v}\to 0}(\delta (2u\bar{v}+\frac{\theta ^{2}}{2}),\tilde{\varphi }) & = & \theta (-u)\int d\phi \varphi (2\sqrt{-u\bar{v}}e_{\theta })=\\
 & = & 2\pi (\theta (-u)\delta _{N}^{(2)}(x),\tilde{\varphi }),\\
\lim _{\bar{v}\to 0}(\bar{v}\delta (2u\bar{v}+\frac{\theta ^{2}}{2})\frac{2}{\theta }e^{i}_{\theta },\tilde{\varphi }) & = & \theta (-u)\bar{v}\int d\phi \frac{e^{i}_{\theta }}{\sqrt{-u\bar{v}}}\varphi (2\sqrt{-u\bar{v}}e_{\theta })=0,\\
\lim _{\bar{v}\to 0}(\frac{1}{\bar{v}}\delta (2u\bar{v}+\frac{\theta ^{2}}{2})\frac{\theta }{2}e^{i}_{\theta },\tilde{\varphi }) & = & \theta (-u)\frac{1}{\bar{v}}\int d\phi \sqrt{-u\bar{v}}e^{i}_{\theta }\varphi (2\sqrt{-u\bar{v}}e_{\theta })=\\
 & = & 2\pi (u\theta (-u)\partial ^{i}\delta _{N}^{(2)}(x),\tilde{\varphi }),
\end{eqnarray*}
where \( \delta _{N}^{(2)}(x) \) denotes the \( \delta  \)-function concentrated
on the north pole (\( \theta =0) \) of the 2-sphere and \( e^{i}_{\theta } \)
the ``radial'' unit-vector, that the asymptotic field-strength is given by
\[
\mathbb {F}=4\pi e(2\theta (-u)\delta ^{(2)}_{N}(x)du+u\theta (-u)dx^{i}\partial _{i}\delta ^{(2)}_{N}(x))\wedge d\bar{v}.\]
 Integration of the flux 2-form over an \( S^{2} \)-section of \( \text {\mcal I}^{+} \)
yields the Bondi-charge
\[
Q^{Bondi}(u)=\frac{1}{4\pi }\int _{S^{2}}*\mathbb {F}=e\theta (-u)\int _{S_{u}^{2}}\delta ^{(2)}_{N}(x)d^{2}\Omega =e\theta (-u).\]
This shows that the charge reaches null infinity at the retarded time \( u=0 \).
The last integration depends only of the ``location'' of the \( S^{2} \)
relative to the point \( u=\theta =0 \) on \( \text {\mcal I}^{+} \) as can
be seen from

\begin{equation}
\label{ascurr}
d*\mathbb {F}|_{\text {\mcal I}^{+}}=4\pi e\delta (u)\delta ^{(2)}_{N}(x)du\wedge d^{2}\Omega .
\end{equation}
The right hand side is actually nothing but the limit of the current to \( \text {\mcal I}^{+} \)
as can be seen from

\begin{eqnarray*}
j=e\delta (px)\delta ^{(2)}(x)(pdx)=4e\delta (u(1+\cos \theta )+v(1-\cos \theta ))\frac{\delta (\sin \theta (v-u))}{2\pi \sin \theta (v-u)} &  & \\
(du(1+\cos \theta )+dv(1-\cos \theta )+(v-u)\sin \theta d\theta )=\left\{ \bar{v}=1/v\right\} = &  & \\
=4e\bar{v}^{3}\delta (\bar{v}u(1+\cos \theta )+(1-\cos \theta ))\frac{\delta (\sin \theta (1-u\bar{v}))}{2\pi \sin \theta (1-u\bar{v})} &  & \\
(du(1+\cos \theta )-\frac{d\bar{v}}{\bar{v}^{2}}(1-\cos \theta )+\bar{v}(1-u\bar{v})\sin \theta d\theta ) &  & \\
\mathbb {J}=\lim _{\bar{v}\to 0}(\Omega ^{-2}j)=e\delta (u)\delta ^{(2)}(x)du, &  & 
\end{eqnarray*}
which is nothing but the dual of (\ref{ascurr}). Finally, let us consider the
limit of the vector potential
\begin{eqnarray}
A=-2e\delta (px)\log \rho \, pdx=-2e\delta (u(1+\cos \theta )+v(1-\cos \theta )) &  & \nonumber \\
\log (\frac{v-u}{2}\sin \theta )\left( du(1+\cos \theta )+dv(1-\cos \theta )+(v-u)\sin \theta d\theta \right) = &  & \nonumber \\
-2e\delta (2u\bar{v}+\frac{\theta ^{2}}{2})(\log \frac{1-u\bar{v}}{2}+\log \sqrt{-4u\bar{v}}-\log \bar{v}) &  & \nonumber \label{veclim} \\
(2\bar{v}du+4ud\bar{v}+(1-u\bar{v})\sqrt{-4u\bar{v}}d\theta )= &  & \nonumber \\
-2e\delta (2u\bar{v}+\frac{\theta ^{2}}{2})(\log ((1-u\bar{v})\sqrt{-u})-\frac{1}{2}\log \bar{v}) &  & \nonumber \\
(2\bar{v}du+4ud\bar{v}+(1-u\bar{v})\sqrt{-4u\bar{v}}d\theta ) &  & 
\end{eqnarray}
As can be seen from (\ref{veclim}) the presence of the \( \log \bar{v} \)
spoil the existence of the limit. However, if one takes a closer look at these
terms one notices that they are actually pure gauge, i.e. of the form \( \log \bar{v}\delta (px)pdx \).
They may therefore be removed by a gauge transformation, which implies that
\[
\lim _{\bar{v}\to 0}(A-e\log \bar{v}\delta (px)pdx)=-4eu\theta (-u)\log (-u)\delta ^{(2)}_{N}(x)d\bar{v}.\]
Since the corresponding (gauge-invariant) field-strength \( F \) exists this
result might actually have been anticipated.

\subsection*{\bit AS-geometry}

There is a striking similarity between the electromagnetic field for an ultrarelativistic
charge and the gravitational field of an ultrarelativistic black hole. The essential
difference is that spacetime is no longer Minkowskian and it is not immediately
clear how to construct a suitable boundary. At this point it is important to
recall that metric (\ref{AS-metric}) being an impulsive pp-wave lies in the
intersection of two larger families of geometries: the Kerr-Schild spacetimes,
which contain pp-waves as subclass on the one hand and impulsive gravitational
waves on the other hand, whose curvature is concentrated on a null hypersurface.
When constructing the asymptotic structure we make use of both of these properties:
The Kerr-Schild form implies that the manifold is equipped with two metrical
structures: a flat one \( \eta _{ab} \) and the physical one \( g_{ab} \);
that the curvature is concentrated on a submanifold of co-dimension one implies
that space is flat almost everywhere. To investigate the asymptotic structure
we compactify the spacetime. We assume that the boundary of the compactified
manifold has the same topology as compactified Minkowski space i.e. \( \mathbb R\times S^{2}\, \cup \, point. \)
Since spacetime is flat almost everywhere we apply the usual conformal rescaling
of flat spacetime to \( g_{ab} \) and show that it has a well-defined (distributional)
meaning on the boundary. The singular nullplane where the curvature is concentrated
cuts \( \text {\mcal I}^{+} \) along a single generator up to the point \( u=0 \).Thus
the conformally transformed physical metric will be smooth everywhere on \( \text {\mcal I}^{+} \)
except for this half generator. As we shall show, on the half-generator the
metric can be given distributional meaning; actually \( g_{ab} \) is continuous.
Although the conformal connection is ill defined all the quantities needed to
define Bondi four momentum do exist. Going out to spatial infinity \( i^{0} \)
the situation is similar: spacelike directions lying in the singular plane trace
out a topological cylinder, i.e. \( \mathbb R\times S^{1} \), on the hyperboloid
\( \mathcal{D} \) in the tangent space at \( i^{0} \) ( \( \theta =\pi /2,\, 0\leq \phi \leq 2\pi  \)).
Once again it can be shown that these contributions of the non-flat part of
the metric vanish in a distributional sense and that the ADM four-momentum is
well defined. 

As already mentioned the metric (\ref{AS-metric}) is flat almost everywhere.
Actually a stronger statement, namely that (\ref{AS-metric}) is asymptotic
to \( \eta _{ab} \) can be proven. To this end let us consider a conformal
factor \( \Omega  \) that gives rise to the standard (Penrose) compactification
of Minkowksi space. Multiplication of (\ref{AS-metric}) with \( \Omega ^{2} \)
gives
\begin{eqnarray*}
\tilde{g}_{ab} & := & \Omega ^{2}g_{ab}=\tilde{\eta }_{ab}+f(\Omega p)_{a}(\Omega p)_{b}=\\
 & = & \tilde{\eta }_{ab}+\delta (\Omega \cdot (px))\Omega (\log (\Omega \rho )-\log \Omega )(\Omega p)_{a}(\Omega p)_{b},
\end{eqnarray*}
where the negative homogeneity of \( \delta  \) has been used in the last equality.
Since \( \Omega p_{a} \), \( \Omega \cdot (px) \) and \( \Omega \cdot \rho  \)
all tend to well-defined (finite) quantities in the limit \( \Omega \to 0 \)
(at the support of \( \delta  \)) it is easy to see that remaining (infinite)
\( \log \Omega  \) cannot outrun \( \Omega  \) . Thus we find
\[
\lim _{\Omega \to 0}\tilde{g}_{ab}=\lim _{\Omega \to 0}\tilde{\eta }_{ab},\]
which explicitly shows that in a suitable framework (i.e. distribution theory)
the AS-geometry tends to the flat metric. Although the metric itself is well-defined
and tends to Minkowski at the boundary the connection is not differentiable.
This is most easily exhibited by taking the \( \bar{v} \)-derivative of 
\begin{eqnarray*}
\Omega ^{2}ds^{2}=dud\bar{v}+\frac{(1-u\bar{v})^{2}}{4}d\Omega ^{2}+\delta (u\bar{v}(1+\cos \theta )+(1-\cos \theta )) &  & \\
(\bar{v}du(1+\cos \theta )-\frac{d\bar{v}}{\bar{v}}(1-\cos \theta -\sin \theta (1-u\bar{v})d\theta )^{2}\frac{\bar{v}}{2}(\log \frac{1-u\bar{v}}{2}\sin \theta -\log \bar{v}). &  & 
\end{eqnarray*}
 Obviously the last factor contains a product of the form \( \bar{v}\log \bar{v} \)
which is not \( \bar{v} \)-differentiable at \( \bar{v}=0 \). However this
term is precisely of the same form as the singular term in the gauge potential
for the electromagnetic shock wave. It is thus reasonable to assume that it
may be eliminated in the same manner by using a singular coordinate transformation.
Although we will not go into the details of this construction, this assumption
is a posteriori justified since the gauge-invariant quantities like the Riemann
and Weyl-tensors have well-defined limits as will be shown in the following. 

The Riemann tensor for an arbitrary pp-wave is given by
\[
R^{ab}\, _{cd}=-2p^{[a}\partial ^{b]}p_{[c}\partial _{d]}f.\]
 Proceeding in an analogous manner as for the Schwarzschild geometry and taking
into account that the wave profile for the AS-geometry is given by \( f=-8\mu \log \rho \delta (px) \),
we obtain
\begin{eqnarray*}
\text {\bf Riemann}=\frac{16\mu \delta (u(1+\cos \theta )+v(1-\cos \theta ))}{(v-u)^{2}\sin ^{2}\theta }[(\sin \theta (\partial _{u}\wedge \partial _{v})- &  & \\
-\frac{1-\cos \theta }{v-u}(\partial _{u}\wedge \partial _{\theta })+\frac{1+\cos \theta }{v-u}(\partial _{v}\wedge \partial _{\theta }))(2\sin \theta (du\wedge dv)+ &  & \\
+(1+\cos \theta )(v-u)(du\wedge d\theta )-(1-\cos \theta )(v-u)(dv\wedge d\theta ))- &  & \\
-((1-\cos \theta )\partial _{u}+(1+\cos \theta )\partial _{v}-\frac{2\sin \theta }{(v-u)}\partial _{\theta })\wedge \partial _{\phi } &  & \\
((1+\cos \theta )du+(1-\cos \theta )dv+\sin \theta (v-u)d\theta )\wedge d\phi ]. &  & 
\end{eqnarray*}
where \( u \) and \( v \) denote the standard (spherical) retarded and advanced
coordinates.\\
\( \bullet  \) \( i^{0} \) and ADM-momentum: For regular coordinates at spatial
infinity \( \bar{u}=-1/v,\bar{v}=-1/u \) the expression for the Riemann tensor
becomes
\begin{eqnarray*}
\text {\bf Riemann}=-(\bar{u}\bar{v})^{3}\frac{16\mu \delta (\bar{u}(1+\cos \theta )+\bar{v}(1-\cos \theta ))}{(\bar{v}-\bar{u})^{2}\sin ^{2}\theta }[(\sin \theta (\partial _{\bar{u}}\wedge \partial _{\bar{v}})+ &  & \\
-\frac{\bar{v}}{\bar{u}}\frac{1-\cos \theta }{\bar{v}-\bar{u}}(\partial _{\bar{v}}\wedge \partial _{\theta })+\frac{\bar{u}}{\bar{v}}\frac{1+\cos \theta }{\bar{v}-\bar{u}}(\partial _{\bar{u}}\wedge \partial _{\theta }))(2\sin \theta (d\bar{u}\wedge d\bar{v})+ &  & \\
+(1+\cos \theta )\frac{\bar{u}}{\bar{v}}(\bar{v}-\bar{u})(d\bar{v}\wedge d\theta )-(1-\cos \theta )\frac{\bar{v}}{\bar{u}}(\bar{v}-\bar{u})(d\bar{u}\wedge d\theta ))- &  & \\
-((1-\cos \theta )\frac{\bar{v}}{\bar{u}}\partial _{\bar{v}}+(1+\cos \theta )\frac{\bar{u}}{\bar{v}}\partial _{\bar{u}}+\frac{2\sin \theta }{(\bar{v}-\bar{u})}\partial _{\theta })\wedge \partial _{\phi } &  & \\
(\frac{\bar{u}}{\bar{v}}(1+\cos \theta )d\bar{v}+\frac{\bar{v}}{\bar{u}}(1-\cos \theta )d\bar{u}-\sin \theta (\bar{v}-\bar{u})d\theta )\wedge d\phi ]. &  & 
\end{eqnarray*}
Taking the direction dependent limit using \( \bar{u}=\lambda U,\, \bar{v}=\lambda V \)
and \( \lambda \to 0 \) we find
\begin{eqnarray*}
\lim _{\lambda \to 0}\Omega ^{-3/2}\text {\bf Riemann}=(-UV)^{3/2}\frac{16\mu \delta (U(1+\cos \theta )+V(1-\cos \theta ))}{(V-U)^{2}\sin ^{2}\theta }[(\sin \theta  &  & \\
(\partial _{U}\wedge \partial _{V})-\frac{V}{U}\frac{1-\cos \theta }{V-U}(\partial _{V}\wedge \partial _{\theta })+\frac{U}{V}\frac{1+\cos \theta }{V-U}(\partial _{U}\wedge \partial _{\theta }))(2\sin \theta (dU\wedge dV)+ &  & \\
+(1+\cos \theta )\frac{U}{V}(V-U)(dV\wedge d\theta )-(1-\cos \theta )\frac{V}{U}(V-U)(dU\wedge d\theta ))- &  & \\
-((1-\cos \theta )\frac{V}{U}\partial _{V}+(1+\cos \theta )\frac{U}{V}\partial _{U}+\frac{2\sin \theta }{V-U}\partial _{\theta })\wedge \partial _{\phi } &  & \\
 &  & \\
(\frac{U}{V}(1+\cos \theta )dV+\frac{V}{U}(1-\cos \theta )dU-\sin \theta (V-U)d\theta )\wedge d\phi ]. &  & 
\end{eqnarray*}
Due to the (infinite) rescaling, i.e. \( \lambda \to 0 \), only the projective
information is physically relevant, which is conveniently encoded in the restriction
to the hyperboloid \( \mathcal{D}:UV=-1 \) 

\begin{eqnarray*}
\mathbb {R}^{ab}\, _{cd} & := & \lim _{\lambda \to 0}\Omega ^{-3/2}R^{ab}\, _{cd}|_{\mathcal{D}}=\frac{16\mu \delta ((1/V)(1+\cos \theta )-V(1-\cos \theta ))}{(V+1/V)^{2}\sin ^{2}\theta }[(\sin \theta +\\
 &  & (\partial _{U}\wedge \partial _{V})^{ab}+V^{2}\frac{1-\cos \theta }{V+1/V}(\partial _{V}\wedge \partial _{\theta })^{ab}-\frac{1}{V^{2}}\frac{1+\cos \theta }{V+1/V}(\partial _{U}\wedge \partial _{\theta })^{ab})\\
 &  & (2\sin \theta (dU\wedge dV)_{cd}+-(1+\cos \theta )\frac{1}{V^{2}}(V+1/V)(dV\wedge d\theta )_{cd}+\\
 &  & +(1-\cos \theta )V^{2}(V+1/V)(dU\wedge d\theta )_{cd})-(((1-\cos \theta )V^{2}\partial _{V}+\\
 &  & (1+\cos \theta )\frac{1}{V^{2}}\partial _{U}-\frac{2\sin \theta }{V+1/V}\partial _{\theta })\wedge \partial _{\phi })^{ab}((\frac{1}{V^{2}}(1+\cos \theta )dV+\\
 &  & V^{2}(1-\cos \theta )dU+\sin \theta (V+1/V)d\theta )\wedge d\phi )_{cd}].
\end{eqnarray*}
As for the Schwarzschild solution the dual of the contraction of \( \mathbb {R}^{ab}\, _{cd} \)
with the bivector potential \( Q^{ab} \) gives the flux 2-form. Its corresponding
integral over an \( S^{2} \)-section of \( \mathcal{D} \) can be expressed
in terms of the Ashtekar-Hansen gravi-electric field \( E^{a}\, _{b}=\eta _{c}\eta ^{d}\mathbb {R}^{ab}\, _{cd} \)

\begin{eqnarray*}
E^{a}\, _{b}=\frac{16\mu \delta ((1/V)(1+\cos \theta )-V(1-\cos \theta ))}{(V+1/V)^{2}\sin ^{2}\theta }(\sin \theta (V\partial _{V}+\frac{1}{V}\partial _{U})^{a}- &  & \\
(1-\frac{V-1/V}{V+1/V}\cos \theta )\partial _{\theta }^{a})(\sin \theta (\frac{1}{V}dV+VdU)_{b}+ &  & \\
(1-\frac{V-1/V}{V+1/V}\cos \theta )\frac{(V+1/V)^{2}}{2}d\theta _{b}) & = & \\
4\mu \frac{\delta (\th \tau -\cos \theta )}{\ch ^{3}\tau }(\partial _{\tau }-\frac{1}{\ch \tau }\partial _{\theta })^{a}(d\tau +\ch \tau d\theta )_{b}. &  & 
\end{eqnarray*}
The ADM-momentum therefore becomes
\begin{eqnarray*}
P^{ADM}\cdot \alpha =-\frac{1}{8\pi }\int _{S^{2}_{\tau }}\, ^{3}\epsilon _{abc}E^{c}\, _{n}\alpha ^{n}=-\frac{\mu }{2\pi }\int d^{2}\Omega \frac{\delta (\th \tau -\cos \theta )}{\ch \tau } &  & \\
(\alpha ^{t}\ch \tau -\sh \tau (\alpha \cdot e_{r})+(\alpha \cdot e_{\theta }))=-\mu (\alpha ^{t}-\alpha ^{z})=\mu (p\cdot \alpha ), &  & 
\end{eqnarray*}
where the integral in the penultimate equality is a formal expression denoting
the evaluation of a (compactly-supported) distribution on the unit test-function.
The calculation is actually independent of the \( S^{2} \) section of \( \mathcal{D} \)
as can be seen from the fact that \( E^{a}\, _{b} \) is conserved, i.e. \( D_{a}E^{a}\, _{b}=0. \)
Thus the ADM-momentum is null and proportional to \( \mu  \).\\
\( \bullet  \) \( \text {\mcal I}^{+} \) and Bondi-momentum: Unlike the Schwarzschild
situation the impulsive gravitaional wave is generated by a null source. Therefore
we will first investigate the behaviour of the latter on the conformal boundary.
This is important, since a failure in the fall-off of the matter distribution
would spoil asymptotic flatness.
\begin{eqnarray*}
T_{ab}=\mu \delta (px)\delta ^{(2)}(x)(pdx)^{2}_{ab}= &  & \\
2\delta (u(1+\cos \theta )+v(1-\cos \theta ))\frac{\delta ((v-u)\sin \theta )}{2\pi (v-u)\sin \theta } &  & \\
\left( du(1+\cos \theta )+dv(1-\cos \theta )+(v-u)d\theta \right) ^{2}_{ab}= &  & \hspace {1cm}\bar{v}=\frac{1}{v}\\
2\bar{v}^{3}\mu \delta (u\bar{v}(1+\cos \theta )+(1-\cos \theta ))\frac{\delta ((1-u\bar{v})\sin \theta )}{2\pi (1-u\bar{v})\sin \theta } &  & \\
(du(1+\cos \theta )-\frac{d\bar{v}}{\bar{v}}(1-\cos \theta )+\frac{1}{\bar{v}}(1-u\bar{v})\sin \theta d\theta )_{ab}^{2} &  & 
\end{eqnarray*}
Taking the limit \( \bar{v}\to 0 \) we find 
\[
\lim _{\bar{v}\to 0}\Omega ^{-2}T_{ab}=\mu \delta (u)\delta ^{(2)}_{N}(x)(du^{2})_{ab}.\]
This result implies that the energy momentum vanishes at the conformal boundary
and therefore does not compromise asymptotic flatness. Moreover, the scaling
behavior is complete agreement with \cite{PenRi}. Having settled this issue,
we begin the calculation of the Bondi-momentum by transforming the expression
for the Riemann-tensor to coordinates regular in a neighborhood of \( \text {\mcal I}^{+} \)
by setting \( \bar{v}=1/v \) 
\begin{eqnarray}
\lefteqn {\text {\bf Riemann}} &  & \hspace {1.3cm}=16\mu \frac{\bar{v}^{3}\delta (\frac{\theta ^{2}}{2}+2u\bar{v})}{\theta ^{2}}[2\theta ^{2}(\partial _{u}\wedge \partial _{\bar{v}})(du\wedge d\bar{v})\nonumber \label{scrRiemann} \\
 &  & -2\theta \bar{v}(\partial _{u}\wedge \partial _{\bar{v}})(du\wedge d\theta )-\frac{\theta ^{3}}{2\bar{v}}(\partial _{u}\wedge \partial _{v})(d\bar{v}\wedge d\theta )+\frac{\theta ^{3}}{\bar{v}}(\partial _{u}\wedge \partial _{\theta })(du\wedge d\bar{v})\nonumber \\
 &  & -\theta ^{2}(\partial _{u}\wedge \partial _{\theta })(du\wedge d\theta )-\frac{\theta ^{4}}{4\bar{v}^{2}}(\partial _{u}\wedge \partial _{\theta })(d\bar{v}\wedge d\theta )+4\theta \bar{v}(\partial _{\bar{v}}\wedge \partial _{\theta })(du\wedge d\bar{v})\nonumber \\
 &  & -4\bar{v}^{2}(\partial _{\bar{v}}\wedge \partial _{\theta })(du\wedge d\theta )-\theta ^{2}(\partial _{\bar{v}}\wedge \partial _{\theta })(d\bar{v}\wedge d\theta )-\theta ^{2}(\partial _{u}\wedge \partial _{\phi })(du\wedge d\phi )\nonumber \\
 &  & +\frac{\theta ^{4}}{4\bar{v}^{2}}(\partial _{u}\wedge \partial _{\phi })(d\bar{v}\wedge d\phi )-\frac{\theta ^{3}}{2\bar{v}}(\partial _{u}\wedge \partial _{\phi })(d\theta \wedge d\phi )+4\bar{v}^{2}(\partial _{\bar{v}}\wedge \partial _{\phi })(du\wedge d\phi )\nonumber \\
 &  & -\theta ^{2}(\partial _{\bar{v}}\wedge \partial _{\phi })(d\bar{v}\wedge d\phi )+2\theta \bar{v}(\partial _{\bar{v}}\wedge \partial _{\phi })(d\theta \wedge d\phi )+4\theta \bar{v}(\partial _{\theta }\wedge \partial _{\phi })(du\wedge d\phi )+\nonumber \\
 &  & -\frac{\theta ^{3}}{\bar{v}}(\partial _{\theta }\wedge \partial _{\phi })(d\bar{v}\wedge d\phi )+2\theta ^{2}(\partial _{\theta }\wedge \partial _{\phi })(d\theta \wedge d\phi )]
\end{eqnarray}
In the above equation we have made use of the fact that the support of the delta-function
concentrates around the north-pole of the \( S^{2} \) in the limit \( \bar{v}\to 0 \).
\begin{eqnarray}
\lefteqn {\mathbb {R}^{ab}\, _{cd}=\lim _{\bar{v}\to 0}(\Omega ^{-3}R^{ab}\, _{cd})=} &  & \hspace {4cm}32\pi \theta (-u)[2\delta _{N}^{(2)}(x)(\partial _{u}\wedge \partial _{\bar{v}})^{ab}(du\wedge d\bar{v})_{cd}+\nonumber \label{scrRiemann1} \\
 &  & +u\partial _{i}\delta _{N}^{(2)}(x)(\partial _{u}\wedge \partial _{\bar{v}})^{ab}(d\bar{v}\wedge dx^{i})_{cd}-2u\partial ^{i}\delta _{N}^{(2)}(x)(\partial _{u}\wedge \partial _{i})^{ab}(du\wedge d\bar{v})_{cd}+\nonumber \\
 &  & +\delta _{N}^{(2)}(x)(\partial _{u}\wedge \partial _{i})^{ab}(du\wedge dx^{i})_{cd}+\delta _{N}^{(2)}(x)(\partial _{\bar{v}}\wedge \partial _{i})^{ab}(d\bar{v}\wedge dx^{i})_{cd}+\nonumber \\
 &  & +u\partial _{j}\delta _{N}^{(2)}(x)(\partial _{u}\wedge \partial _{i})^{ab}(dx^{j}\wedge dx^{i})_{cd}+2u\partial ^{i}\delta ^{(2)}_{N}(x)(\partial _{i}\wedge \partial _{j})^{ab}(d\bar{v}\wedge dx^{j})_{cd}-\nonumber \\
 &  & -\delta _{N}^{(2)}(x)(\partial _{i}\wedge \partial _{j})^{ab}(dx^{i}\wedge dx^{j})_{cd}+u^{2}(\partial ^{i}\partial _{j}-\frac{1}{2}\delta ^{i}_{j}\partial ^{2})\delta _{N}^{(2)}(x)\nonumber \\
 &  & (\partial _{u}\wedge \partial _{i})^{ab}(d\bar{v}\wedge dx^{j})_{cd}]
\end{eqnarray}
where \( x^{i} \) refer to the standard two-dimensional Cartesian coordinates
associated with circular polar coordinates. Let us emphasize that in the limit
expression (\ref{scrRiemann1}) not all the terms of (\ref{scrRiemann}) have
finite limits by themself but fortunately these infinities cancel out in the
final result. The relevant integrals for deriving these identities arise from
calculations similar to
\begin{eqnarray*}
\lefteqn {\lim _{\bar{v}\to 0}(\frac{\theta ^{2}}{\bar{v}^{2}}\delta (\frac{\theta ^{2}}{2}+2u\bar{v})(e^{i}_{\theta }e^{j}_{\theta }-e^{i}_{\phi }e^{j}_{\phi }),\tilde{\varphi })=} &  & \\
 &  & =\theta (-u)\int d\phi \frac{-4u\bar{v}}{\bar{v}^{2}}(e^{i}_{\theta }e^{j}_{\theta }-e^{i}_{\phi }e^{j}_{\phi })\varphi (\sqrt{-4u\bar{v}}e_{\theta })=\\
 &  & =8u^{2}\theta (-u)\int d\phi (e^{i}_{\theta }e^{j}_{\theta }-e^{i}_{\phi }e^{j}_{\phi })e^{k}_{\theta }e^{l}_{\theta }\partial _{k}\partial _{l}\varphi (0)=\\
 &  & =4u^{2}\theta (-u)\pi (-\delta ^{ij}\partial ^{2}\varphi (0)+2\partial ^{i}\partial ^{j}\varphi (0)).
\end{eqnarray*}
 This result shows that, as in the case of the energy momentum tensor the (rescaled)
Weyl-tensor has a well-defined distributional limit which is completely concentrated
on the generator of \( \text {\mcal I}^{+} \) corresponding to the north pole
of the 2-sphere for negative retarded times. Taking the integral of the flux
2-form \( \frac{1}{2}\, ^{4}\epsilon _{abcd}(\mathbb {R}^{ab}\, _{cd}Q^{cd}\frac{1}{2}) \)
finally yields the Bondi four-momentum
\begin{eqnarray*}
P^{Bondi}\cdot \alpha  & = & \frac{1}{8\pi }\int _{S^{2}_{u}}\frac{1}{2}\, ^{4}\epsilon _{abcd}(\mathbb {R}^{ab}\, _{cd}Q^{cd}\frac{1}{2})=\\
 &  & -\frac{32\mu \pi }{8\pi }\theta (-u)\int d^{2}\Omega \frac{1}{4}\delta ^{(2)}_{N}(x)(\alpha ^{t}-(a\cdot e_{r}))=\mu \theta (-u)(p\cdot \alpha ).
\end{eqnarray*}
 This confirms our previously obtained result that the Bondi-momentum is null
and equal to ADM up to the instant of retarded time \( u \), where both particle
and wave reach infinity and drops to zero. All energy escapes to \( \text {\mcal I}^{+} \)
which is flat thereafter.

\section*{\large\bit Conclusion}

In this paper we have taken some effort to put on a stricter mathematical footing
a result which on physical grounds is intuitively expected: The boosted Schwarzschild
black hole looks more and more like a gravitational wave as the boost parameter
tends to the velocity of light. Therefore, if in the limit process the energy
is kept constant, the resulting gravitational wave should have a lightlike ADM
and Bondi four-momentum. However, since there are theorems which exclude the
existence of light-like total four-momenta in general relativity \cite{AshHor}
on the basis of regularity assumptions for spacetime, the limiting metric has
to violate these. In fact the curvature is of distributional nature concentrated
on a null hypersurface. Nevertheless, we have shown that it is possible to construct
a conformal boundary for this metric in a distributional sense. This implies
that some quantities are only defined over the space of testfunctions (forms)
at spatial and null infinity. Since the integrand for the Bondi-momentum is
proportional to \( \theta (-u) \), the Bondi-news square would behave like
\( \delta (u) \). Therefore the news-function itself is not defined within
the framework of classical distribution theory.

Bicak and Schmidt have analyzed boost symmetric spacetimes \cite{BiSchm}. Specific
examples represent the field of two uniformly accelerated particles. These spacetimes
have a regular \( \text {\mcal I} \) except at points where the particles reach
infinity. Similarly, the trajectory of the ultrarelativistic ``particle''
meets \( \text {\mcal I} \) in a point. However, we have shown that the associated
stress-energy tensor vanishes at the conformal boundary. The reason for this
is that the mass parameter of the Schwarzschild black hole tends to zero in
the ultra-relativistic limit. Therefore, in our case there are no punctures
in \( \text {\mcal I} \), which has been shown using distributional techniques.
However, it should be kept in mind that the framework of distributions does
in general not allow to assign point values.

Among the class of pp-waves the AS metric can be considered to be on the ``edge''
of asymptotic flatness. It is just the right combination \( \delta \cdot \log  \)
terms in the wave profile which renders a finite non-zero four-momenta: Widening
the impulsive profile to would make the four-momenta undefined, while replacing
the \( \log  \)-term by a (stronger) falloff, gives a zero total momentum (see
\cite{BeiChr})

Although our construction refers to a specific spacetime it seems reasonable
to believe that distributional techniques may be applicable to asymptopia in
a more general context. Mathematically this due to the fact that in the asymptotically
flat situation gravity becomes weak ``far away'' from the sources and may
be treated in a linear approximation. However, classical distribution theory
is tailor-made to deal with non-smooth, linear situations, thereby transferring
the burden of differentiability from the boundary manifold to the space of (generalized)
functions thereon. 

\vspace{2cm}
\noindent \textbf{Acknowledgments:} One of us (PCA) thanks A. Ashtekar and B.
Schmidt for clarifying discussions during a program at the ITP of the UCSB.
This program was supported by the NSF grant no. PHY 94-07194. Partial support
by the Fundacion Federico is also acknowledged.

\end{document}